\documentclass[10pt,conference,a4paper]{IEEEtran}
\IEEEoverridecommandlockouts

\newif\ifanon
\anonfalse    

\usepackage{cite}
\usepackage{amsmath,amssymb,amsfonts}
\usepackage{algorithmic}
\usepackage{graphicx}
\usepackage{multirow}
\usepackage{braket}
\usepackage{textcomp}
\usepackage{xcolor}
\usepackage{tikz}
\usetikzlibrary{arrows.meta, positioning, shapes.geometric, fit, backgrounds}
\usepackage{booktabs}
\def\BibTeX{{\rm B\kern-.05em{\sc i\kern-.025em b}\kern-.08em
    T\kern-.1667em\lower.7ex\hbox{E}\kern-.125emX}}
\usepackage[colorlinks=true, allcolors=blue]{hyperref}
\usepackage{url}

\begin{document}

\title{Learning Trotter Orderings for Heisenberg Hamiltonians with a Ranking Transformer}

\ifanon
  \author{\IEEEauthorblockN{Anonymous Author(s)}
  \IEEEauthorblockA{Submission under double-blind review}}
\else
  \author{
  \IEEEauthorblockN{Shamminuj Aktar, Reuben Tate, and Stephan Eidenbenz}
  \IEEEauthorblockA{\textit{Computing and Artificial Intelligence Division (CAI-3)} \\
  \textit{Los Alamos National Laboratory, Los Alamos, NM, USA}\\
  \{saktar, rtate, eidenben\}@lanl.gov}
  }
\fi

\maketitle

\begin{abstract}

    Trotterization approximates quantum time evolution by sequentially applying Hamiltonian terms. Because noncommuting terms introduce ordering-dependent errors, selecting an optimal term ordering is a combinatorial problem over a factorial search space. Prior approaches rely on fixed heuristics or on selection among predefined structured orderings, both of which require simulating candidates before choosing among them.
    Using 1D and 2D Heisenberg-style Hamiltonians as a guiding example, we instead learn the ordering directly with a physics-aware ranking transformer that assigns a scalar score to each Hamiltonian term and predicts an ordering by sorting these scores. Physical structure enters through commutator-biased attention and an anticommutation-weighted ranking loss, and the models are trained on simulated-annealing (SA) reference orderings. We train separate models for first- and second-order Trotterization, each jointly on chains up to $14$ qubits and lattices up to $12$ qubits, and evaluate them on unseen chains with 16--20 qubits and lattices with $16$ and $20$ qubits.
    Predicted orderings are evaluated by the median gap in simulation fidelity to the SA reference. At first order, the model reaches a pooled gap below $10^{-4}$ on chains, $0.0144$ and $0.0088$ on triangular lattices, and $0.0948$ and $0.0823$ on rectangular lattices at $16$ and $20$ qubits; at second order, the corresponding gaps are $0.0114$, $0.0364$ and $0.0244$, and $0.1348$ and $0.1239$. The prediction exceeds the SA reference on $34$\% of first-order and $8$\% of second-order chain instances and can match its fidelity even when the two sequences differ because commuting terms may be rearranged without changing the Trotter unitary.
   The learned models generalize to larger systems across all three geometries, performing best on chains and triangular lattices, and produce an ordering in one forward pass without candidate enumeration, simulated annealing, or fidelity evaluation.

\end{abstract}


\section{Introduction}
\label{sec:intro}

    Trotterization~\cite{trotter1959product,lloyd1996universal,suzuki1991general} is a standard method for approximating the time evolution of a Hamiltonian system and is an important building block in many fault-tolerant quantum algorithms, including ground-state energy estimation through quantum phase estimation~\cite{reiher2017elucidating}, quantum linear-system solvers~\cite{PhysRevLett.103.150502}, and the simulation of lattice Hamiltonians~\cite{haah2018quantum,georgescu2014quantum}. The method decomposes a Hamiltonian into local terms and applies their exponentials sequentially. For a Hamiltonian with $K$ terms, these terms may be applied in any of $K!$ possible orderings. When the terms do not commute, different orderings generally produce different simulation errors~\cite{babbush2015error,childs2021trotter}. Selecting an effective ordering is therefore a combinatorial optimization problem over a factorial search space.
    
  Existing ordering methods rely primarily on hand-designed heuristics, such as grouping commuting terms or sorting terms by coefficient magnitude~\cite{tranter2019ordering,tate2026commutation}, while other work optimizes the ordering for circuit depth rather than simulation fidelity~\cite{schmitz2024graphoptimization}. No single heuristic performs best across all Hamiltonian geometries and Trotter configurations~\cite{tate2026commutation}, so selecting among them still requires simulating multiple candidate orderings.
    
    We address this problem with a transformer that learns to generate an ordering directly from the Hamiltonian. We study one- and two-dimensional Heisenberg-style Hamiltonians and evaluate whether the learned models generalize to system sizes larger than those used in training. Heisenberg models are widely used benchmarks for quantum simulation and describe low-energy magnetic phenomena such as magnetic order, spin correlations, and phase transitions~\cite{auerbach2012interacting}. The systems considered here are motivated by quantum magnets such as Cs$_2$CoCl$_4$~\cite{PhysRevLett.127.037201} and KYbSe$_2$~\cite{scheie2021witnessing}. Our approach differs from the prior learned method of Aktar et al.~\cite{aktar2026structure}, which formulates ordering selection as classification over a fixed set of $24$ structured candidates and was demonstrated only on one-dimensional systems. We formulate the task as candidate-free ranking. A physics-aware transformer assigns a scalar score to each Hamiltonian term, and sorting these scores produces the predicted ordering in a single forward pass. The models are trained using simulated-annealing (SA) reference orderings.
    
    The main contributions of this work are:
    \begin{itemize}
    \item We formulate Trotter term ordering as a candidate-free ranking problem that produces an ordering directly from the Hamiltonian in a single forward pass.
    \item We develop a physics-aware transformer~\cite{vaswani2017attention} that incorporates anticommutation structure through both commutator-biased attention and an anticommutation-weighted ranking loss.
    \item We evaluate the method across three Hamiltonian geometries and two Trotter orders, demonstrating strong size generalization on one-dimensional chains and triangular lattices while identifying rectangular lattices as a challenging regime.
    \end{itemize}

    \begin{figure*}[t!]
      \centering
      \includegraphics[width=0.95\textwidth]{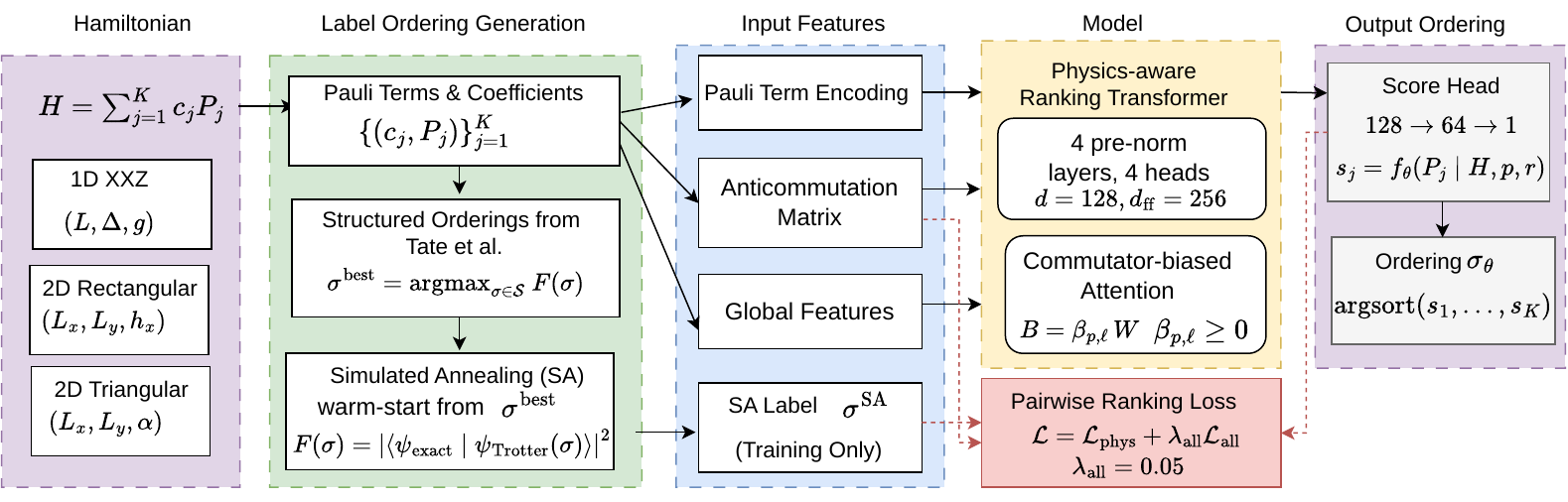}
       \vspace{-5pt}
   \caption{Overview of the method. \emph{Hamiltonian:} the 1D XXZ chain and 2D rectangular and triangular lattices. \emph{Label Generation:} the highest-fidelity structured candidate $\sigma^{\mathrm{best}}$ seeds simulated annealing, ensuring that the SA reference $\sigma^{\mathrm{SA}}$ is no worse than the best structured baseline. \emph{Input Features:} per-term Pauli encodings, global context, and the anticommutation matrix $W$, all read from the Hamiltonian. \emph{Model Learning:} $W$ biases attention and weights the pairwise ranking loss. \emph{Model Prediction:} scalar term scores are sorted to obtain $\sigma_\theta$. Red dashed lines indicate training-only paths; at inference, the model produces an ordering in one forward pass without candidate enumeration or fidelity evaluation.}
      \label{fig:overview}
    \vspace{-7pt}
    \end{figure*}
    
    \section{Background and Related Work}
    \label{sec:background}
    
    \subsection{Trotterization and Term Ordering}
    
   A product formula approximates $e^{-iHt}$ for a Hamiltonian $H=\sum_{j=1}^{K} c_jP_j$, where $c_j$ is the coefficient and $P_j$ is the Pauli term, by dividing the evolution time $t$ into $r$ steps and applying single-term exponentials within each step according to a permutation $\sigma\in\Pi_K$~\cite{lloyd1996universal,suzuki1991general}. The first-order formula applies each term once per step, while the second-order formula applies the ordering followed by its reverse, with each term evolved for half a step. Following Tate et al.~\cite{tate2026commutation}, we measure ordering quality using the simulation fidelity $F(\sigma)$, defined as the squared overlap between the exactly evolved and Trotterized states. Both formulas are exact when all terms commute; otherwise, $F(\sigma)$ generally depends on $\sigma$. Most theoretical analyses derive worst-case bounds from nested commutators and operator norms~\cite{childs2021trotter}, but these bounds do not generally identify the best ordering for a specific Hamiltonian and initial state.
    
    \subsection{Ordering Heuristics}
    
    Tranter et al.~\cite{tranter2019ordering} proposed several ordering rules for electronic-structure simulation, including \texttt{depleteGroups}, \texttt{equaliseGroups}, coefficient-magnitude ordering, and lexicographic ordering. Tate et al.~\cite{tate2026commutation} studied commutation-based strategies for Heisenberg-type Hamiltonians by partitioning terms into mutually commuting groups through colorings of the commutation graph and then ordering the groups. Their results show that the best strategy depends on the Hamiltonian geometry and Trotter configuration, and identifying it requires evaluating the candidates directly. Commutation-graph colorings perform strongly on triangular lattices but are less effective on rectangular lattices under second-order Trotterization.
    
    \subsection{Learned Orderings}
    
    Aktar et al.~\cite{aktar2026structure} formulated ordering selection as classification over $24$ structured candidates. The approach generalizes to chains larger than those used during training, but it can only return an ordering from the predefined candidate set. The candidates must also be constructed and simulated for each new Hamiltonian geometry, and the method was demonstrated only on one-dimensional systems. Our physics-aware transformer instead produces a candidate-free ordering directly from the Hamiltonian in a single forward pass.
    
    \begin{figure*}[t]
    \centering
    \includegraphics[width=0.95\linewidth]{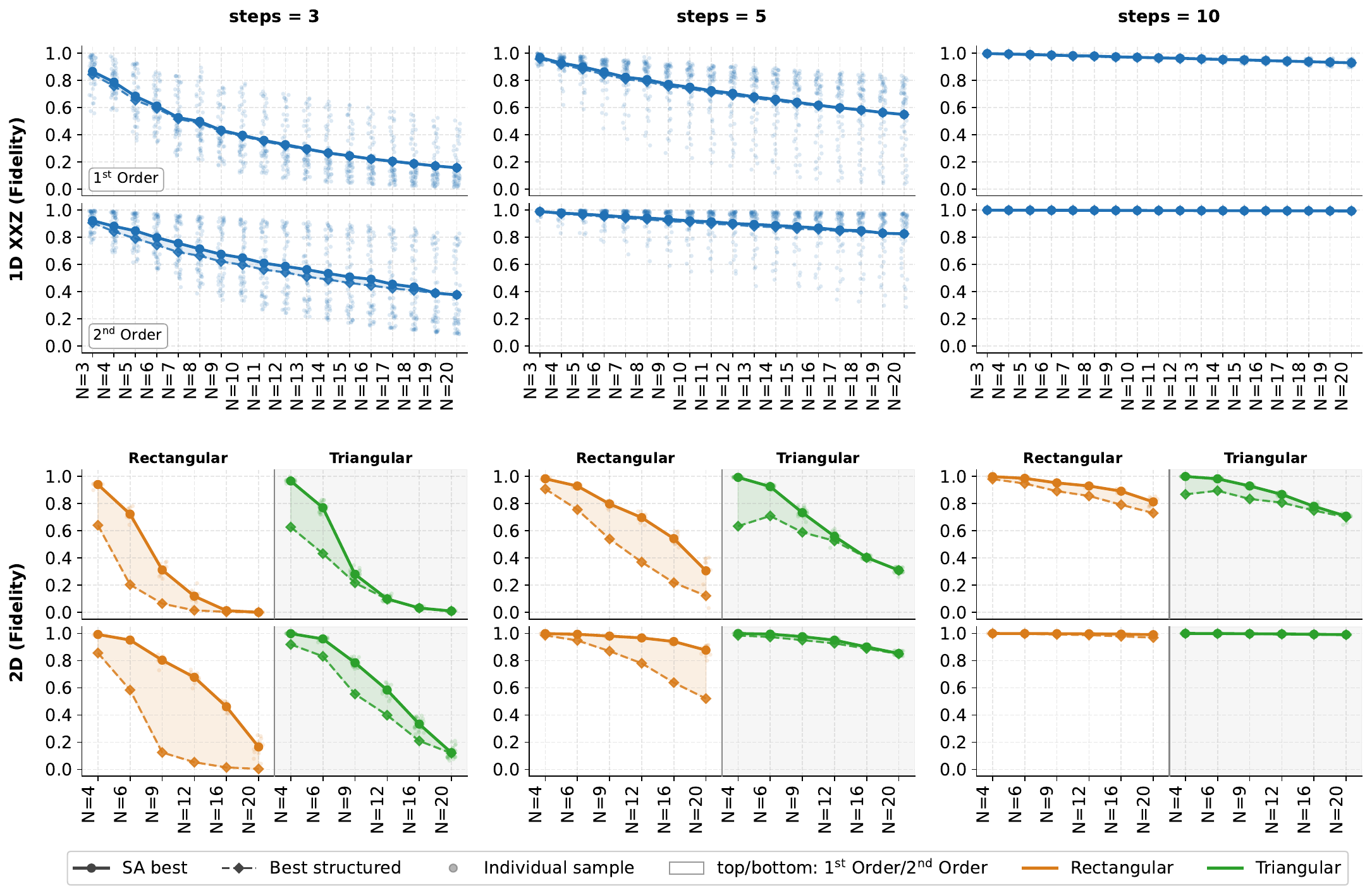}
    \vspace{-4pt}
   \caption{Comparison of the SA reference and the best structured candidate. Solid lines show the SA-reference fidelity, dashed lines the best structured-candidate fidelity, and the shaded region the improvement from SA; faint markers denote individual runs. The top row shows the 1D XXZ chain, and the bottom row the rectangular (left) and triangular (right) lattice families. Columns correspond to Trotter step counts, and each panel is divided into first-order (top) and second-order (bottom) strips. The improvement is small for chains, where structured candidates are strong, and substantially larger for both lattice families.}
    \label{fig:labelq}
     \vspace{-7pt}
    \end{figure*}

\section{Methodology}
\label{sec:method}

    \subsection{Problem Formulation}
    \label{sec:method-task}
    
   Given a Hamiltonian $H=\sum_{j=1}^{K} c_jP_j$, together with a product-formula order $p$ and step count $r$, we represent an ordering by a permutation $\sigma\in\Pi_K$, where $\Pi_K$ is the set of all $K!$ permutations of the terms. The optimal ordering is $\sigma^\star = \operatorname*{arg\,max}_{\sigma\in\Pi_K}F(\sigma)$. Direct optimization is computationally infeasible as the search space grows with the number of Pauli terms. For example, a 20-site XXZ chain has $K=77$ terms and therefore $77!$ possible orderings, each requiring simulation of the Trotterized evolution and comparison with the exact reference evolution. We instead formulate learning Trotter orderings as a ranking problem. The model assigns a scalar score $s_j=f_\theta(P_j\mid H,p,r)$ to each term and sorts the scores to obtain $\sigma_\theta=\operatorname{argsort}(s_1,\ldots,s_K)$. This produces an ordering directly from the Hamiltonian without evaluating fidelity at inference. Because the ordering is constructed by sorting scalar scores, the model represents relative precedence between terms rather than absolute sequence positions.
    
    \subsection{Simulated-Annealing (SA) Labels}
    \label{sec:method-labels}
    
    For each Hamiltonian and Trotter configuration, we evaluate the structured candidate set $\mathcal{S}$ of Tate et al.~\cite{tate2026commutation}, obtained from colorings of the commutation graph and all permutations of the resulting color groups, giving $24$ candidates for chains and $6$ for lattices. SA is initialized from the best candidate $\sigma^{\mathrm{best}} = \operatorname*{arg\,max}_{\sigma\in\mathcal{S}}F(\sigma)$, rather than from a random permutation, so the retained ordering is at least as good as $\sigma^{\mathrm{best}}$.
    
    SA uses three moves: swapping two terms, swapping adjacent terms, and removing a term and reinserting it at a random position. The move distribution favors random swaps for chains and is more even across the three types for lattices, where reinsertion is more effective on the larger term sets. A proposal is accepted whenever it improves the fidelity and otherwise with probability $\exp(\Delta F/T)$. The temperature is cooled geometrically from $T_{\mathrm{start}}=0.05$ to $T_{\mathrm{end}}=10^{-4}$, and the best ordering encountered is retained as the SA reference $\sigma^{\mathrm{SA}}$. The SA budget is given in Section~\ref{sec:setup-data}. The improvement over the seed, $\Delta F_{\mathrm{SA}} = F(\sigma^{\mathrm{SA}}) - F(\sigma^{\mathrm{best}}) \geq 0$, measures the fidelity recovered by SA. Figure~\ref{fig:labelq} shows this gain is small for chains, where the structured candidates are already strong, and much larger for both lattice families. SA therefore gives a strong reference, but it does not certify a global optimum, so a prediction may occasionally exceed it.

    \subsection{Pairwise Ranking Loss}
    \label{sec:ranking-loss}
    
    Ordering-dependent Trotter error arises from noncommuting terms because two commuting exponentials may be exchanged without changing the Trotterized unitary. We build the training objective around this observation by penalizing an incorrectly ordered pair in proportion to the strength of its anticommutation relation. We encode this information in the matrix
    \begin{equation}
    W_{ij} =
    \begin{cases}
    |c_i c_j| & \text{if } P_i \text{ and } P_j \text{ anticommute},\\
    0 & \text{otherwise},
    \end{cases}
    \label{eq:wcomm}
    \end{equation}
    which is normalized by its largest entry. Commuting pairs therefore receive zero weight.
    
    Let $\pi(j)$ denote the position of term $j$ in $\sigma^{\mathrm{SA}}$. Because the predicted ordering is obtained by sorting scores in ascending order, every pair for which the SA reference places $i$ before $j$ should satisfy $s_i<s_j$. We penalize violations using
    \begin{equation}
    \mathcal{L}_{\mathrm{phys}}
    = \frac{\sum\nolimits_{\pi(i)<\pi(j)} W_{ij}\,\zeta(s_i-s_j)}
           {\sum\nolimits_{\pi(i)<\pi(j)} W_{ij}} ,
    \label{eq:loss-phys}
    \end{equation}
    where $\zeta(x)=\log(1+e^x)$ is the softplus function. The penalty approaches zero when the precedence relation is satisfied with a sufficiently large margin and grows approximately linearly when it is strongly violated. Normalizing by the total pair weight makes the loss comparable across system sizes.
    
    Because $\mathcal{L}_{\mathrm{phys}}$ leaves commuting pairs unconstrained, we add an unweighted loss $\mathcal{L}_{\mathrm{all}}$, defined as in Eq.~\eqref{eq:loss-phys} with $W_{ij}$ replaced by unity. The full objective is
    \begin{equation}
    \mathcal{L} = \mathcal{L}_{\mathrm{phys}} + \lambda_{\mathrm{all}}\mathcal{L}_{\mathrm{all}},
    \qquad \lambda_{\mathrm{all}}=0.05 .
    \label{eq:loss-total}
    \end{equation}
    The gradient is therefore dominated by physically consequential anticommuting pairs, while the remaining pairs are weakly anchored to the reference. Since the physics-weighted objective constrains only anticommuting pairs, rearranging commuting terms incurs no penalty under $\mathcal{L}_{\mathrm{phys}}$.

    \subsection{Model Architecture}
    \label{sec:method-arch}
    
    The model is a transformer encoder~\cite{vaswani2017attention} operating on the unordered set of Pauli terms and producing one scalar score per term, as shown in Fig.~\ref{fig:overview}. Term representations are formed by concatenating the continuous features described in Section~\ref{sec:setup-enc} with learned embeddings of the categorical features and projecting the result to dimension $d=128$. The encoder contains four pre-normalized layers, four attention heads, a feed-forward width of $256$, and dropout of $0.1$. No positional encoding is used, so the model receives no information about an initial term ordering. The physics prior enters through the attention mechanism. We add the bias $B = \beta_{p,\ell} W$ to the attention logits of every head, where $W$ is the matrix defined in Eq.~\eqref{eq:wcomm} and $\beta_{p,\ell}$ is a learned nonnegative scalar associated with product-formula order $p$ and Hamiltonian geometry $\ell$. A positive $\beta_{p,\ell}$ encourages anticommuting terms to attend to one another. Learning this weight separately across regimes allows the influence of the physics prior to vary with the Trotter order and Hamiltonian geometry.
    
   The global context of each instance, comprising the step count, product-formula order, Hamiltonian geometry, and the scalar summaries described in Section~\ref{sec:setup-enc}, is mapped by a separate encoder to a $d$-dimensional vector that is added to every term representation. This allows a single model to serve all step counts and Hamiltonian geometries within a fixed product-formula order. Chains have no lattice geometry, so we gate the geometry pathway off for them and train its parameters on lattice instances alone. 
   
   A per-term head with dimensions $128\to64\to1$ produces the scalar scores, which are sorted to obtain the predicted ordering. The model contains approximately $583{,}000$ parameters. We train one model for each product-formula order, jointly across all Hamiltonian geometries and step counts.
\section{Experimental Setup}
\label{sec:exp_setup}

\subsection{Hamiltonian Families}
\label{sec:setup-hams}

We use the three Heisenberg-style Hamiltonian families and the evaluation setting of Tate et al.~\cite{tate2026commutation}. The first is the transverse-field XXZ chain, motivated by the quantum magnet Cs$_2$CoCl$_4$~\cite{PhysRevLett.127.037201},
\begin{equation}
H_{\mathrm{1D}} = \sum_{\langle i,j\rangle}
\left( X_iX_j + Y_iY_j + \Delta Z_iZ_j \right)
+ g\sum_i X_i,
\label{eq:h1d}
\end{equation}
in dimensionless units. We use $\Delta\in\{0.12,0.25\}$ and $g\in\{0.1,0.2,\ldots,2.5\}$, giving $50$ Hamiltonians per system size. The chains have open boundaries and lengths $L=3,\ldots,20$, so each instance contains $K=4L-3$ Pauli terms.

Writing $S_{ij}=X_iX_j+Y_iY_j+Z_iZ_j$ for the isotropic Heisenberg coupling, the first two-dimensional family is the $J_1$--$J_2$ antiferromagnet on a triangular lattice,
\begin{equation}
H_{\mathrm{tri}} = \sum_{\langle i,j\rangle}S_{ij}
+ \alpha \sum_{\langle\langle i,j\rangle\rangle}S_{ij},
\label{eq:htri}
\end{equation}
which is motivated by the spin-liquid candidate KYbSe$_2$~\cite{scheie2021witnessing}. The second sum runs over next-nearest-neighbor pairs, and the frustration parameter $\alpha=J_2/J_1$ is swept over $27$ values in $[0,0.5]$. The second lattice family is the rectangular lattice
\begin{equation}
H_{\mathrm{rect}} = \sum_{\langle i,j\rangle}S_{ij} + h_x\sum_i X_i,
\label{eq:hrect}
\end{equation}
with $h_x$ swept over $12$ values in $[0,3.0]$. Both lattice families use open boundaries and snake qubit indexing on $L_x\times L_y$ grids containing $4$--$20$ qubits, giving $27$ triangular and $12$ rectangular Hamiltonians per system size. All simulations begin from the N\'eel state $\ket{0101\ldots}$. The total evolution time is $T=5.0$ for chains and $T=1.0$ for lattices. The shorter lattice evolution time keeps these more highly connected systems in a regime where term ordering continues to produce measurable fidelity differences. We consider first- and second-order product formulas with step counts $r\in\{3,5,10\}$.

\subsection{Dataset and Splits}
\label{sec:setup-data}

Each sample consists of a Hamiltonian, a Trotter configuration, and its SA reference ordering. We split the data by system size to evaluate extrapolation beyond the training range. Training uses chains with $L\leq14$ and lattices with $q\leq12$. Validation uses chains with $L=15$ and lattices with $q=16$. The reported out-of-training-size evaluation includes chains with $L=16$--$20$ and lattices with $q\in\{16,20\}$.

A separate model is trained for each product-formula order. For each order, the dataset contains $3402$ samples: $2268$ for training, $267$ for validation, and $984$ for evaluation. The validation and evaluation sets overlap because the $117$ lattice samples at $q=16$ serve both roles. Only two lattice sizes lie beyond the training range, so $q=16$ is used for validation and is also reported as an intermediate out-of-training-size result. It is not used for parameter updates, but because validation is performed on it, it is not an independent test set. We therefore report the $q=16$ and $q=20$ results separately, with $q=20$ serving as the fully held-out lattice test size. The chain size $L=15$ is used only for validation and is excluded from all reported test results.

The SA budget decreases with system size because each iteration requires a Trotterized simulation. 1D chains receive $2000$ iterations up to $L=16$, $1000$ iterations at $L=17$ and $L=18$, and $500$ iterations at $L=19$ and $L=20$. Lattices receive $2000$ iterations up to $q=16$ and $500$ iterations at $q=20$. The effect of this reduction depends on the quality of the initial structured candidate. For chains, the structured candidates already lie close to the SA reference, as shown in Fig.~\ref{fig:labelq}, so the reduced budget has little effect. For lattices, the structured seeds are weaker and SA recovers substantially more fidelity. Nevertheless, the reduced SA budget at $q=20$ still improves on the structured seed for most instances, yielding a useful reference at the largest system size.

\begin{figure*}[t]
\centering
\includegraphics[width=0.95\linewidth]{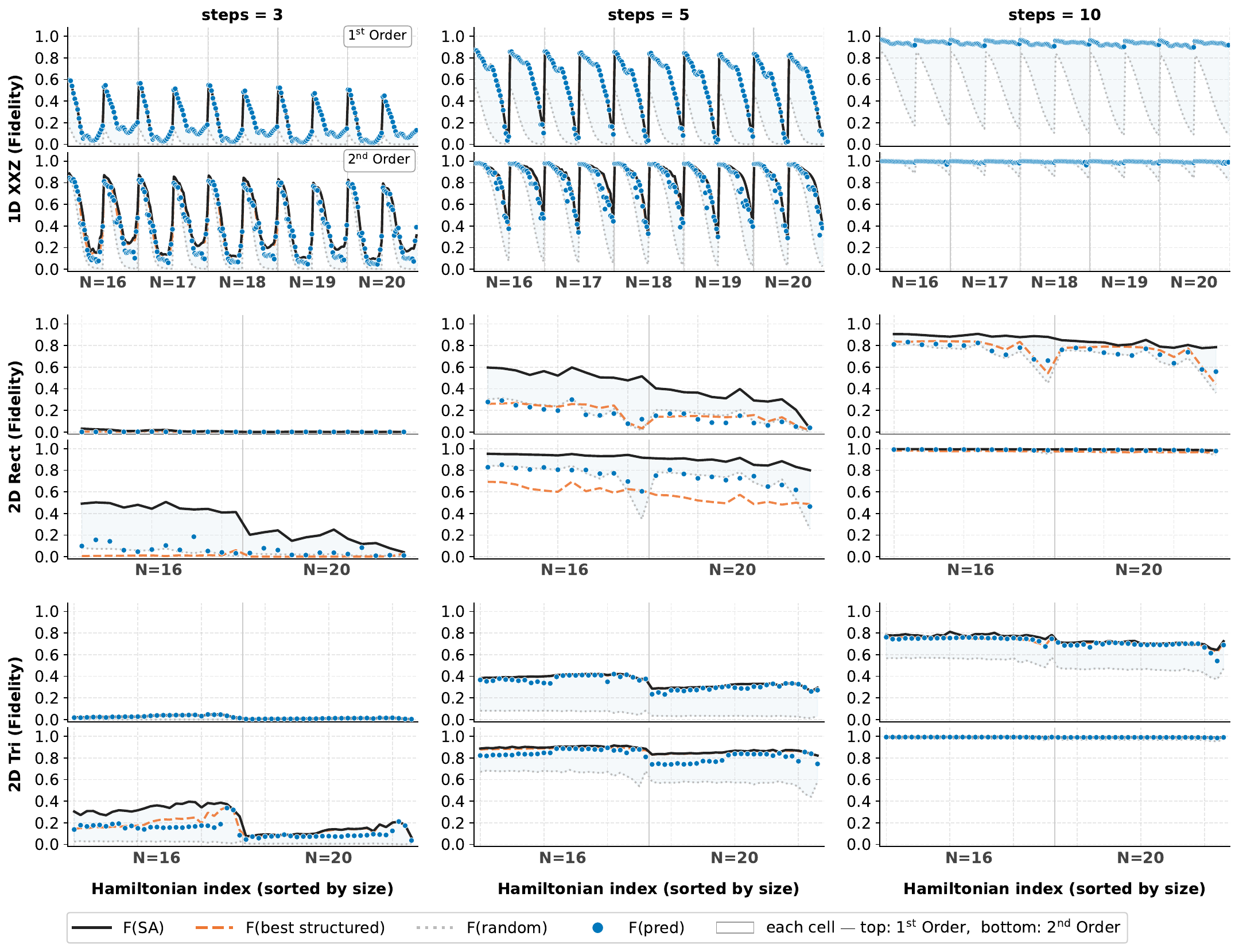}
\vspace{-5pt}
\caption{Per-Hamiltonian fidelity on systems larger than those used for training. Each marker represents one evaluated instance. Instances are sorted by system size, with vertical rules separating size groups. Lines show the SA reference (solid), the best structured candidate (dashed), and the mean random-ordering fidelity (dotted), while the shaded region spans the range between the random ordering and the SA reference. Rows correspond to Hamiltonian families and columns to Trotter step counts; each panel is divided into first-order (top) and second-order (bottom) strips.}
\label{fig:lines}
 \vspace{-8pt}
\end{figure*}

\subsection{Per-Term Encoding}
\label{sec:setup-enc}

Each Pauli term is represented by eleven continuous features and three categorical embeddings, as shown in Fig.~\ref{fig:overview}. The continuous features include the log-magnitude of the coefficient and three commutation features derived from the normalized matrix $W$ in Eq.~\eqref{eq:wcomm}: the total anticommutation weight $\sum_iW_{ij}$, the maximum pairwise weight $\max_iW_{ij}$, and the fraction of the remaining $K-1$ terms that anticommute with term $j$. Four locality features encode the logarithm of the qubit span, the minimum site index normalized by $n_q-1$, the parity of that index, and an indicator for whether the term acts on the first qubit. Three geometric features encode the normalized lattice position of the bond and the logarithm of its Euclidean length. The categorical embeddings specify the operator type ($X$, $Y$, $Z$, $XX$, $YY$, or $ZZ$), whether the term is one- or two-body, and the bond direction.

Bond coordinates are recovered from snake-indexed site indices only for the rectangular family. Triangular-lattice instances receive zeros in all geometric feature slots, while the geometry pathway is gated off for chains. The rectangular lattice is therefore the only family for which explicit geometric information reaches the model.

Each sample also includes the matrix $W$, which is used in both the attention bias and the ranking loss, and a global-context vector. The global context contains embeddings of the step count, product-formula order, and Hamiltonian geometry, together with six scalar descriptors: coefficient spread, coefficient entropy, two-body fraction, log aspect ratio, mean coordination number, and boundary fraction.

\subsection{Training and Evaluation}
\label{sec:setup-train}

Models are trained using AdamW with learning rate $3\times10^{-4}$, weight decay $10^{-4}$, a cosine learning-rate schedule over $300$ epochs, batch size $16$, and gradient clipping at unit norm. Training stops early when the validation Kendall tau~\cite{kendall1938new} between the predicted and reference orderings does not improve for $60$ epochs, and the checkpoint with the highest validation agreement is retained. Each configuration is trained using three random seeds, and reported predicted fidelities are averaged across seeds.

We evaluate a predicted ordering using the fidelity it achieves and report its gap to the SA reference,
\begin{equation}
\Delta F = F(\sigma^{\mathrm{SA}}) - F(\sigma_\theta).
\label{eq:gap}
\end{equation}
Here $\sigma_\theta$ is the predicted ordering (Section~\ref{sec:method-task}); $\Delta F=0$ means that the prediction matches the reference fidelity, while $\Delta F<0$ means that it exceeds the reference. Because the gap distribution is skewed, we report the median. For context, we also report the fidelity of the best structured ordering, $F(\mathrm{best})$, and the mean fidelity over random permutations, $F(\mathrm{rand})$. Together, these baselines indicate how strongly the simulation fidelity depends on term ordering for a given Hamiltonian.

\begin{figure*}[t]
\centering
\includegraphics[width=0.95\linewidth]{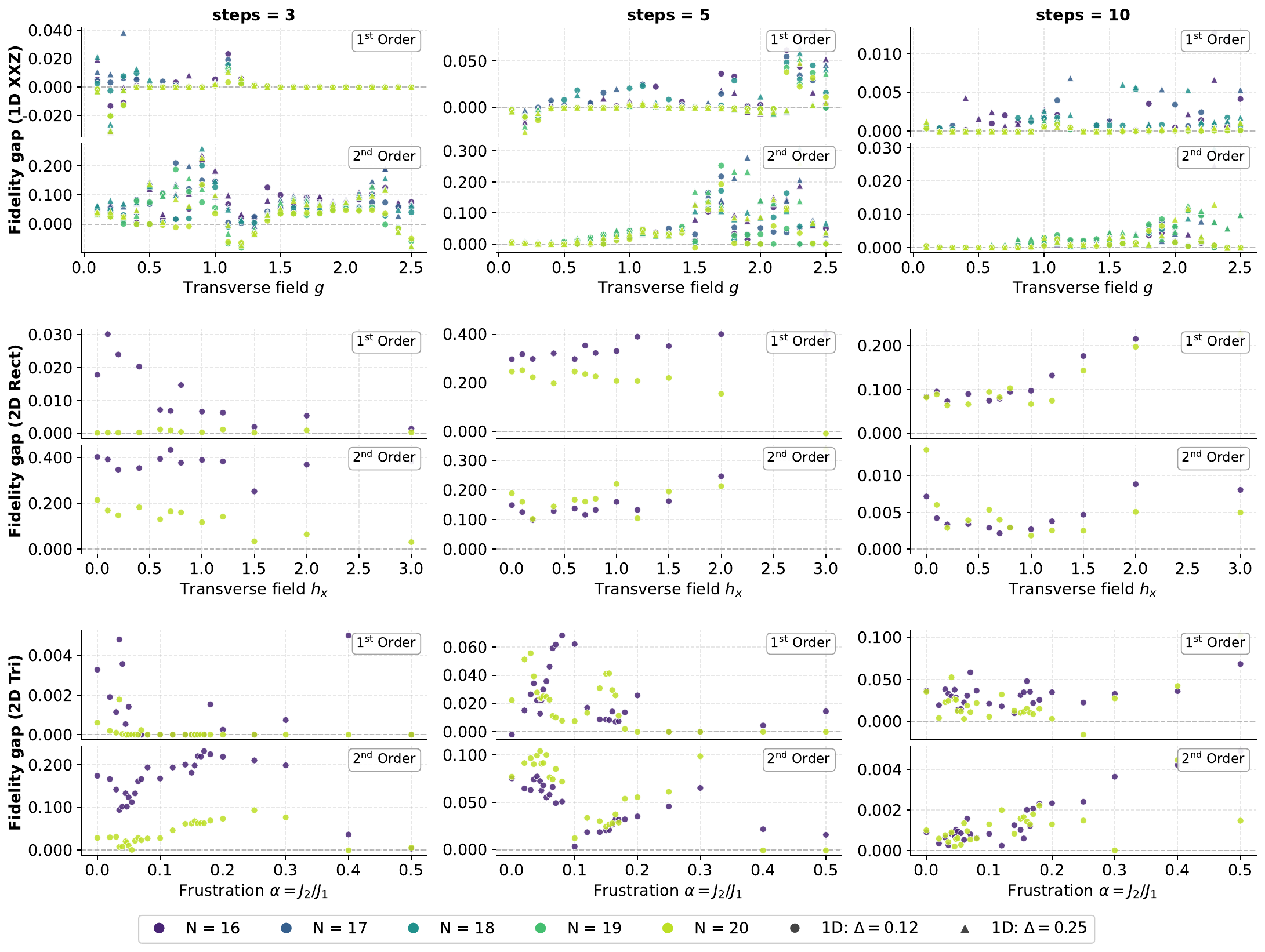}
\vspace{-5pt}
\caption{Fidelity gap to the SA reference as a function of the Hamiltonian parameter on systems larger than those used for training. Rows correspond to Hamiltonian families, with sweeps over $g$ for chains, $\alpha=J_2/J_1$ for triangular lattices, and $h_x$ for rectangular lattices; columns correspond to Trotter step counts. Each panel is divided into first-order (top) and second-order (bottom) strips. Each marker shows the median gap over instances sharing the same parameter value, with color indicating system size. For chains, circles and triangles distinguish the two anisotropies $\Delta$. Points below the dashed zero line indicate predictions that exceed the SA reference, which is possible because SA does not certify a global optimum. The gap remains small for chains and triangular lattices across their parameter ranges and is largest for rectangular lattices.}
\label{fig:param}
 \vspace{-5pt}
\end{figure*}

\section{Experimental Results} 
\label{sec:result}

    We evaluate the trained models on systems larger than those used in training (Section~\ref{sec:setup-data}). Table~\ref{tab:results} summarizes the results, Fig.~\ref{fig:lines} compares the predicted fidelity with the SA reference, best structured candidate, and mean random ordering for each instance, and Fig.~\ref{fig:param} shows how the gap varies across each parameter sweep. Predicted fidelities are averaged over three seeds, and reported fidelities and gaps are medians over instances.
    \begin{table}[t]
    \centering
    \caption{Median fidelities and gaps on larger systems outside training sets. Chain results are pooled over $L=16$--$20$. The quantity $\Delta F$ is the fidelity gap to the SA reference. The final column gives the fraction of instances for which the prediction outperforms the best structured candidate. $F(\sigma_\theta)$ needs no simulation; $F(\mathrm{best})$ needs 24 (chains) or 6 (lattices), and SA 500--2000.}
    \label{tab:results}
    \footnotesize
    \setlength{\tabcolsep}{4pt}
    \begin{tabular}{@{}lccccc@{}}
    \toprule
    & $\Delta F$ & $F(\sigma_\theta)$ & $F(\mathrm{best})$ & $F(\mathrm{rand})$ & $>$best \\
    \midrule
    \multicolumn{6}{@{}l}{\textit{First order}} \\
    1D                 & $<10^{-4}$ & 0.670 & 0.672 & 0.087 & 43\% \\
    2D Tri $q=16$      & 0.0144 & 0.376 & 0.400 & 0.081 & 25\% \\
    2D Tri $q=20$      & 0.0088 & 0.292 & 0.305 & 0.033 & 16\% \\
    2D Rect $q=16$     & 0.0948 & 0.203 & 0.249 & 0.243 & 14\% \\
    2D Rect $q=20$     & 0.0823 & 0.090 & 0.141 & 0.156 & 17\% \\
    \midrule
    \multicolumn{6}{@{}l}{\textit{Second order}} \\
    1D                 & 0.0114 & 0.878 & 0.911 & 0.791 & 14\% \\
    2D Tri $q=16$      & 0.0364 & 0.847 & 0.888 & 0.673 & 38\% \\
    2D Tri $q=20$      & 0.0244 & 0.777 & 0.849 & 0.572 & 36\% \\
    2D Rect $q=16$     & 0.1348 & 0.805 & 0.627 & 0.810 & 94\% \\
    2D Rect $q=20$     & 0.1239 & 0.726 & 0.508 & 0.742 & 94\% \\
    \bottomrule
    \end{tabular}
    \vspace{-5pt}
    \end{table}
    
    \subsection{One-Dimensional Chains}
    \label{sec:res-1d}
    
   For first-order Trotterization, the prediction closely matches the SA reference at every unseen chain length, with a pooled median gap below $10^{-4}$. The large separation between the SA-reference and random-ordering fidelities in Fig.~\ref{fig:lines} shows that term ordering substantially affects these systems. The model recovers essentially all of the SA-reference fidelity on chains of length up to $L=20$, containing $77$ Pauli terms. The model achieves this even without reproducing the SA reference. Commuting terms can be exchanged without changing the Trotterized unitary, so the SA reference is only one of potentially many orderings with the same fidelity. The physics-weighted ranking loss in Section~\ref{sec:ranking-loss} reflects this freedom by not penalizing the relative arrangement of commuting pairs. The predicted ordering exceeds the SA reference on 34\% of first-order instances, increasing from 21\% at $L=16$ to 45\% at $L=20$. This is possible because SA does not certify a global optimum. The effect occurs most frequently at the largest sizes, where the reduced SA budget can leave the SA reference equal to the best structured candidate. Second-order Trotterization is more difficult. The pooled median gap is $0.0114$, and the prediction exceeds the SA reference on only 8\% of instances. 
    
    \subsection{Triangular Lattices}
    \label{sec:res-tri}
    
    The models generalize well to triangular lattices larger than those used during training. The median gap decreases from $0.0144$ at $q=16$ to $0.0088$ at $q=20$ for first order and from $0.0364$ to $0.0244$ for second order. Performance therefore does not degrade across the two evaluated sizes beyond the training range. The prediction exceeds the mean random-ordering fidelity on every evaluated triangular-lattice instance and outperforms the best structured candidate on 36\% of second-order instances at $q=20$. This performance is obtained without explicit triangular-lattice geometry. As described in Section~\ref{sec:setup-enc}, triangular-lattice instances do not contain geometric features, so their predictions rely only on coefficient, commutation, and locality information.

     \subsection{Rectangular Lattices}
    \label{sec:res-rect}
    
    Rectangular lattices are the most difficult family and the only one for which the prediction remains substantially below the SA reference. The median gaps are several times larger than the triangular-lattice gaps at both Trotter orders.
    
   The quality of the SA labels does not explain this behavior. Fig.~\ref{fig:labelq} shows that SA improves the structured seeds on rectangular lattices. At $q=16$, it increases the first-order fidelity from $0.249$ to $0.539$ and the second-order fidelity from $0.627$ to $0.943$, improving on the seed in $33$ of $36$ first-order instances and all $36$ second-order instances. A similar pattern appears at $q=20$. SA therefore identifies much stronger orderings, but the model does not fully recover them. At $q=16$ and first order, the median predicted fidelity of $0.203$ is also below the mean random-ordering fidelity of $0.243$.
    
   Geometric features help on rectangular lattices, lowering the pooled gap from $0.0937$ to $0.0893$ at first order and from $0.2363$ to $0.1325$ at second order, yet the gap remains far above the triangular gaps. Rectangular lattices are the only family for which explicit geometry reaches the model, since triangular instances receive zeros in these slots. The model nevertheless performs better on triangular lattices, which suggests that the current geometric encoding, while useful, does not fully capture the structure most relevant to strong rectangular orderings.
    
    A distinguishing feature of this family is the weakness of the available structured candidates. Consistent with Tate et al.~\cite{tate2026commutation}, the best structured candidate performs worse than the mean random ordering in $31$ of $36$ second-order instances at $q=16$ and $33$ of $36$ at $q=20$. SA must therefore move far from its seed, and initializing it from a broader heuristic set~\cite{tranter2019ordering} or the best random ordering could yield stronger references. The model nevertheless recovers part of this improvement. At second order, it outperforms the best structured candidate on 94\% of instances at both sizes while still trailing the SA reference by a substantial margin. First-order performance is weaker, with the prediction exceeding the best structured candidate on only 14\%--17\% of instances.

    \subsection{Dependence on Hamiltonian Parameters}
    \label{sec:res-sweep}
    
    Figure~\ref{fig:param} shows the fidelity gap against the swept parameter of each family: the transverse field $g$ for chains, the frustration $\alpha=J_2/J_1$ for triangular lattices, and the transverse field $h_x$ for rectangular lattices. Across all three families, the gap varies little over the parameter range, so the difficulty of a family is largely a property of its geometry rather than of a particular physical regime. For chains, the gap stays near zero across the full field sweep at first order, with a modest rise near $g=1.5$--$2.3$ at second order for step counts of $5$ and $10$. For triangular lattices, it remains small across the full frustration range, including the low-$\alpha$ region below $\alpha=0.2$. For rectangular lattices, it stays large throughout the field sweep, confirming that the difficulty of this family is not confined to a narrow regime. Points below the zero line, most visible for chains at first order, mark instances where the prediction exceeds the SA reference.

    \subsection{Trotter Order and Step Count} 
    \label{sec:res-order}
    
   Two effects appear across all three Hamiltonian families. First, some parameter settings are nearly insensitive to ordering. At $r=3$ and first order, the SA-reference fidelity is $0.017$ on triangular lattices and $0.0015$ on rectangular lattices, with mean random-ordering fidelities near zero, while at $r=10$ and second order every method exceeds $0.99$ on all three families. As Fig.~\ref{fig:lines} shows, the SA reference, best structured, and random fidelities collapse together in both regimes, so the band between them nearly vanishes and a small gap does not necessarily indicate an effective ordering.
    
    Outside them, the gap is larger for second-order than for first-order Trotterization in every family (Table~\ref{tab:results}), and the prediction exceeds the SA reference on 34\% of first-order chain instances but only 8\% of second-order instances. One possible reason is that the second-order formula applies the sequence forward and then in reverse, so a term's contribution may depend on its absolute position as well as its precedence relations, and sorting scalar scores represents precedence directly but position only indirectly.
    
    The second effect is that the second-order gap decreases as the number of Trotter steps increases. As $r$ grows from $3$ to $10$, it falls from $0.0564$ to $0.0008$ on chains, from $0.0750$ to $0.0010$ on triangular lattices, and from $0.2333$ to $0.0039$ on rectangular lattices, since more steps shorten the interval per step, reducing the effect of noncommutativity and making ordering less consequential. For first-order lattice simulations this trend does not hold, because the small-$r$ cases lie in the nearly flat, low-fidelity regime above. 
    
    Finally, an ablation shows that the two physics components provide complementary benefits. Removing the attention bias widens the first-order rectangular gap from $0.0893$ to $0.1028$, while replacing the anticommutation-weighted loss with uniform pair weights widens the triangular gap from $0.0107$ to $0.0373$. The full model performs best on every Hamiltonian family at both Trotter orders.

\section{Conclusion and Future Work}
\label{sec:conclusion}

We introduced a physics-aware ranking transformer that predicts a Trotter ordering directly from Hamiltonian structure by scoring each Pauli term and sorting the scores. The model produces an ordering in a single forward pass without enumerating structured candidates, running simulated annealing, or evaluating candidate fidelities at inference. Trained on chains with up to $14$ qubits and lattices with up to $12$, the models generalize to larger systems at both Trotter orders, achieving near-SA-reference fidelity on chains and triangular lattices. The predicted ordering can differ from the SA reference while attaining the same fidelity because commuting terms may be rearranged without changing the Trotterized unitary. 

Rectangular lattices remain the most challenging family. The model stays below the SA reference and drops below the mean random ordering at first order, so performance is not yet uniform across geometries. Since the SA reference orderings are not guaranteed to be optimal, the model is only as good as the labels it trains on, and these are weakest on rectangular lattices where SA starts from poor structured seeds. The evaluation also spans two lattice sizes beyond the training range and is limited to Heisenberg-style Hamiltonians with state fidelity as the objective.

These observations point to several directions for future work. Triangular lattices are handled well without explicit geometric features, whereas rectangular lattices remain difficult despite receiving them, suggesting that the current encoding does not fully capture the relevant lattice structure. Architectures that preserve geometry more directly may improve performance in this regime. Stronger references may be obtained by seeding simulated annealing from a broader heuristic set, improving supervision on the hardest families. The ranking formulation can also be extended to other Hamiltonian families and to objectives that account for circuit depth and gate count alongside state fidelity. All labels use the N\'eel state, so transfer to other initial states remains to be tested.

\ifanon\else
\section*{Acknowledgements}
The research presented in this article was supported by the NNSA's Advanced Simulation and Computing Beyond Moore's Law program at Los Alamos National Laboratory. This material is based upon work supported by the U.S. Department of Energy, Office of Science, National Quantum Information Science Research Centers, Quantum Science Center. This work has been assigned LANL technical report number LA-UR-26-25821.
\fi

\bibliographystyle{IEEEtran}
\bibliography{main}

@article{trotter1959product,
  author    = {Trotter, Hale F.},
  title     = {On the product of semi-groups of operators},
  journal   = {Proceedings of the American Mathematical Society},
  volume    = {10},
  number    = {4},
  pages     = {545--551},
  year      = {1959},
  doi       = {10.2307/2033649}
}

@article{suzuki1991general,
  author    = {Suzuki, Masuo},
  title     = {General theory of fractal path integrals with applications to
               many-body theories and statistical physics},
  journal   = {Journal of Mathematical Physics},
  volume    = {32},
  number    = {2},
  pages     = {400--407},
  year      = {1991},
  doi       = {10.1063/1.529425}
}

@article{lloyd1996universal,
  author    = {Lloyd, Seth},
  title     = {Universal quantum simulators},
  journal   = {Science},
  volume    = {273},
  number    = {5278},
  pages     = {1073--1078},
  year      = {1996},
  doi       = {10.1126/science.273.5278.1073}
}

@article{childs2021trotter,
  author    = {Childs, Andrew M. and Su, Yuan and Tran, Minh C.
               and Wiebe, Nathan and Zhu, Shuchen},
  title     = {Theory of {T}rotter Error with Commutator Scaling},
  journal   = {Physical Review X},
  volume    = {11},
  number    = {1},
  pages     = {011020},
  year      = {2021},
  doi       = {10.1103/PhysRevX.11.011020}
}

@article{babbush2015error,
  author    = {Babbush, Ryan and McClean, Jarrod and Wecker, Dave
               and Aspuru-Guzik, Al{\'a}n and Wiebe, Nathan},
  title     = {Chemical basis of {T}rotter-{S}uzuki errors in quantum
               chemistry simulation},
  journal   = {Physical Review A},
  volume    = {91},
  number    = {2},
  pages     = {022311},
  year      = {2015},
  doi       = {10.1103/PhysRevA.91.022311}
}

@article{tranter2019ordering,
  author    = {Tranter, Andrew and Love, Peter J. and Mintert, Florian
               and Wiebe, Nathan and Coveney, Peter V.},
  title     = {Ordering of {T}rotterization: Impact on Errors in Quantum
               Simulation of Electronic Structure},
  journal   = {Entropy},
  volume    = {21},
  number    = {12},
  pages     = {1218},
  year      = {2019},
  doi       = {10.3390/e21121218}
}

@article{schmitz2024graphoptimization,
  author    = {Schmitz, Albert T. and Sawaya, Nicolas P. D.
               and Johri, Sonika and Matsuura, A. Y.},
  title     = {Graph optimization perspective for low-depth
               {T}rotter-{S}uzuki decomposition},
  journal   = {Physical Review A},
  volume    = {109},
  number    = {4},
  pages     = {042418},
  year      = {2024},
  publisher = {American Physical Society},
  doi       = {10.1103/PhysRevA.109.042418}
}

@misc{tate2026commutation,
  author        = {Tate, Reuben and Aktar, Shamminuj and Eidenbenz, Stephan},
  title         = {An Analysis of Commutation-Based {T}rotter Ordering Strategies
                   on {H}eisenberg-Style {H}amiltonians},
  year          = {2026},
  eprint        = {2604.23138},
  archivePrefix = {arXiv},
  primaryClass  = {quant-ph}
}

@inproceedings{vaswani2017attention,
  author    = {Vaswani, Ashish and Shazeer, Noam and Parmar, Niki
               and Uszkoreit, Jakob and Jones, Llion and Gomez, Aidan N
               and Kaiser, {\L}ukasz and Polosukhin, Illia},
  title     = {Attention is All you Need},
  booktitle = {Advances in Neural Information Processing Systems},
  editor    = {Guyon, I. and Von Luxburg, U. and Bengio, S. and Wallach, H.
               and Fergus, R. and Vishwanathan, S. and Garnett, R.},
  publisher = {Curran Associates, Inc.},
  volume    = {30},
  year      = {2017},
  url       = {https://proceedings.neurips.cc/paper_files/paper/2017/file/3f5ee243547dee91fbd053c1c4a845aa-Paper.pdf}
}

@article{georgescu2014quantum,
  author    = {Georgescu, I. M. and Ashhab, S. and Nori, Franco},
  title     = {Quantum simulation},
  journal   = {Reviews of Modern Physics},
  volume    = {86},
  number    = {1},
  pages     = {153--185},
  year      = {2014},
  doi       = {10.1103/RevModPhys.86.153}
}

@article{haah2018quantum,
  author  = {Haah, Jeongwan and Hastings, Matthew B. and Kothari, Robin
             and Low, Guang Hao},
  title   = {Quantum Algorithm for Simulating Real Time Evolution of
             Lattice {H}amiltonians},
  journal = {SIAM Journal on Computing},
  volume  = {52},
  number  = {6},
  pages   = {FOCS18-250--FOCS18-284},
  year    = {2023},
  doi     = {10.1137/18M1231511}
}

@article{PhysRevLett.127.037201,
  author    = {Laurell, Pontus and Scheie, Allen and Mukherjee, Chiron J.
               and Koza, Michael M. and Enderle, Mechtild
               and Tylczynski, Zbigniew and Okamoto, Satoshi
               and Coldea, Radu and Tennant, D. Alan and Alvarez, Gonzalo},
  title     = {Quantifying and Controlling Entanglement in the Quantum
               Magnet {$\mathrm{Cs}_2\mathrm{CoCl}_4$}},
  journal   = {Physical Review Letters},
  volume    = {127},
  number    = {3},
  pages     = {037201},
  year      = {2021},
  doi       = {10.1103/PhysRevLett.127.037201}
}

@book{auerbach2012interacting,
  author    = {Auerbach, Assa},
  title     = {Interacting Electrons and Quantum Magnetism},
  publisher = {Springer Science \& Business Media},
  year      = {2012},
  doi       = {10.1007/978-1-4612-0869-3}
}

@inproceedings{aktar2026structure,
  title     = {Structure-Aware Transformers for Learning Near-Optimal
               Trotter Orderings with System-Size Generalization in 1D
               Heisenberg Hamiltonians},
  author    = {Aktar, Shamminuj and Tate, Reuben and Eidenbenz, Stephan},
  booktitle = {2026 IEEE International Conference on Quantum Computing
               and Engineering (QCE)},
  year      = {2026},
  doi       = {10.1109/QCE68830.2026.00304}
}

@article{scheie2021witnessing,
  author        = {Scheie, A. O. and Ghioldi, E. A. and Xing, J.
                   and Paddison, J. A. M. and Sherman, N. E. and Dupont, M.
                   and Sanjeewa, L. D. and Lee, S. and Woods, A. J.
                   and Abernathy, D. and others},
  title         = {Witnessing quantum criticality and entanglement in the
                   triangular antiferromagnet {KYbSe$_2$}},
  year          = {2021},
  eprint        = {2109.11527},
  archivePrefix = {arXiv},
  primaryClass  = {cond-mat.str-el}
}

@article{reiher2017elucidating,
  author    = {Reiher, Markus and Wiebe, Nathan and Svore, Krysta M.
               and Wecker, Dave and Troyer, Matthias},
  title     = {Elucidating reaction mechanisms on quantum computers},
  journal   = {Proceedings of the National Academy of Sciences},
  volume    = {114},
  number    = {29},
  pages     = {7555--7560},
  year      = {2017},
  doi       = {10.1073/pnas.1619152114}
}

@article{PhysRevLett.103.150502,
  title = {Quantum Algorithm for Linear Systems of Equations},
  author = {Harrow, Aram W. and Hassidim, Avinatan and Lloyd, Seth},
  journal = {Phys. Rev. Lett.},
  volume = {103},
  issue = {15},
  pages = {150502},
  numpages = {4},
  year = {2009},
  month = {Oct},
  publisher = {American Physical Society},
  doi = {10.1103/PhysRevLett.103.150502}
}

@article{kendall1938new,
    author = {KENDALL, M. G.},
    title = {A NEW MEASURE OF RANK CORRELATION},
    journal = {Biometrika},
    volume = {30},
    number = {1-2},
    pages = {81-93},
    year = {1938},
    month = {06},
    issn = {0006-3444},
    doi = {10.1093/biomet/30.1-2.81},
    url = {https://doi.org/10.1093/biomet/30.1-2.81},
    eprint = {https://academic.oup.com/biomet/article-pdf/30/1-2/81/423380/30-1-2-81.pdf},
}

\end{document}